\documentclass[aps,prl,twocolumn,superscriptaddress]{revtex4-2}
\usepackage{bm}
\usepackage{physics}
\usepackage{anyfontsize}
\usepackage{amsmath}
\usepackage{amsthm}
\usepackage{color}
\usepackage{comment} 
\usepackage{amssymb}

\usepackage{mathtools}
\usepackage{newtxtext}
\usepackage{graphicx}
\usepackage[colorlinks=true,allcolors=blue]{hyperref}

\usepackage{hyperref} 
\usepackage{cleveref} 
\newcommand{\sectionprl}[1]{{\par\it #1.---}}

\usepackage{hyperref} 
\usepackage{cleveref}
\usepackage{tikz}
\usepackage{tikz-3dplot}

\usepackage[normalem]{ulem}
\usepackage[dvipsnames]{xcolor}

\begin{document}
\title{Speed limits for information acquisition in stochastic learning dynamics}

\author{Shuta Kobayashi}
\affiliation{Department of Physics, Kyoto University, Kyoto 606-8502, Japan}

\author{Andreas Dechant}
\affiliation{Department of Physics, Kyoto University, Kyoto 606-8502, Japan}

\begin{abstract}
Neural networks acquire internal representations through learning.
In this work, we formulate stochastic gradient descent (SGD) as a Markovian stochastic process and derive a Fisher-information flow speed limit that bounds the rate at which trainable parameters can acquire information about data-generating parameters.
The resulting inequality decomposes the upper bound on the information flow into drift and noise contributions, thereby quantifying the roles of deterministic learning forces and SGD-induced fluctuations from an information-theoretic perspective.
We verify the bound in analytically tractable basis-function linear regression, where the information budget predicted by the bound reproduces the ordering and characteristic time scales with which different data-generating parameters are encoded in the learned parameters.
These results establish Fisher-information speed limits as a quantitative framework for understanding which properties of the data-generating process are acquired, in what order, and how rapidly they can be encoded in the trainable parameters.
\end{abstract}

\maketitle

\sectionprl{Introduction}\label{sec:intro}
Neural networks underpin modern artificial intelligence, including large language models, yet how information is acquired and processed during training remains an area of active research.
Stochastic gradient descent (SGD) and its variants are widely used for training~\cite{Bottou2010-kh}, and their random mini-batch updates allow learning to be described as a stochastic dynamical process~\cite{Li2017-tu,Yaida2018-qg,li2019stochastic,Ali2020-dm,Mandt2017-fj}.
This stochastic viewpoint provides a natural connection to information theory and stochastic thermodynamics, where information flow in stochastic systems has been extensively studied~\cite{Schreiber2000-nb,Horowitz2014-tt,Ito2013-ys,Goldt2017-hg,Goldt2017-lg,Hartich2014-zm,Hartich2016-zq,matsumoto2018role}.
Related questions arise in cellular sensing and synaptic plasticity, where fluctuating internal states encode information about external signals and their statistics~\cite{Lang2014-qz,Sartori2014-lg,Karbowski2021-ns}.
These connections motivate a broader question: how do stochastic dynamics constrain the rate at which an adaptive system acquires information about distinct properties of its environment?

In neural networks, information-theoretic studies have primarily focused on how information is transformed across network layers~\cite{Tishby2000-nl,tishby2015deep,Shwartz-Ziv2017-va,Saxe2019-eb,pmlr-v97-goldfeld19a}.
More recently, Fisher information has been used to quantify the transmission of parameter-relevant information through trained neural networks~\cite{Weimar2025-rl}.
Complementary studies have investigated how training data shape learned parameters, either through their mutual information with the training set~\cite{Goldt2017-lg,Pensia2018-ah,Futami2023-ok,Parsi2024-gy} or through the alignment of learned weight vectors with teacher or target directions~\cite{Saad1995-hs,saxe2014exact,Goldt2020-iw,JMLR:v22:20-1288,Abbe2022-ii,JMLR:v25:23-1543}.
However, the former captures aggregate statistical dependence without identifying which properties of the data-generating process are encoded, while the latter often relies on restrictive assumptions about network architectures and data distributions.
Consequently, a general understanding of which properties of the data-generating process are acquired by trainable parameters, and on what time scales, remains incomplete.

In the context of stochastic dynamical systems, various speed limit theorems have been established in recent years~\cite{Shiraishi2018-cz,Ito2020-es,Vo2020-cg,Van-Vu2023-ty,Karbowski2024-jb,Nishiyama2026-gd}.
These are upper bounds on the the rate at which, for example,  the  probability  distribution  can  change,  which  are formulated using characteristic features of the system, such as its overall transition rate or its irreversibility. Such speed limits allow us to understand intuitively which features of a stochastic dynamical system allow it to support fast and reliable operations.
Applied to SGD, they motivate a complementary question: which features of the stochastic learning dynamics constrain the rate of information  acquisition, and what can these constraints reveal about the training process?

\begin{figure}[t]
    \centering
    \includegraphics[width=1.0\linewidth]{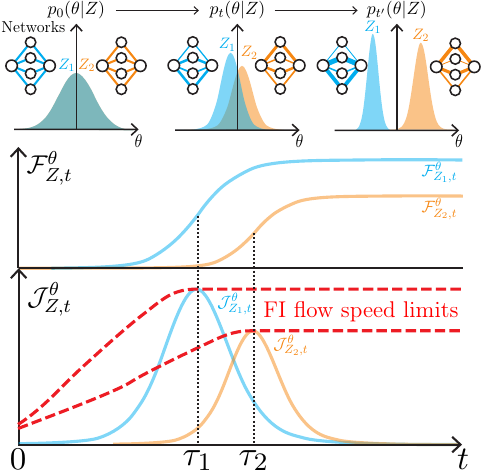}
    \caption{Schematic illustration of Fisher-information acquisition during stochastic learning.
    Top: evolution of the network parameter distribution $p_t(\theta|Z)$ as learning proceeds from left to right. Initially, the distribution is independent of the data-generating parameter $Z$; during learning, information about $Z$ is encoded in the parameters $\theta$, making the distributions associated with $Z_1$ and $Z_2$ increasingly distinguishable.
    Middle: the corresponding Fisher information $\mathcal{F}^{\theta}_{Z,t}$ [Eq.~\eqref{eq:fisher-information}] accumulated about $Z_1$ and $Z_2$, increases as learning proceeds.
    Bottom: the Fisher-information flow $\mathcal{J}^{\theta}_{Z,t}$ [Eq.~\eqref{eq:fi-flow}], which quantifies the instantaneous rate of information acquisition and peak at the characteristic times $\tau_1$ and $\tau_2$. The red dashed curves indicate the corresponding Fisher-information flow speed limits.
    }
    \label{fig:fi-flow}
\end{figure}

Here, we consider data generated from a probability distribution characterized by data-generating parameters $Z$.
As learning proceeds, the distribution $p_t(\theta|Z)$ of the trainable parameters $\theta$ become sensitive to $Z$, and the trainable parameters thereby acquire statistically accessible information about the data-generating process (see Fig.~\ref{fig:fi-flow}).
We quantify this acquired information using the Fisher information
\begin{gather} \label{eq:fisher-information}
    \mathcal{F}^{\theta}_{Z,t} \coloneqq \langle\left|\nabla_Z \log p_t(\theta|Z)\right|^2\rangle_t .
\end{gather}
Through the Cramér--Rao inequality, Fisher information bounds the uncertainty of estimating $Z$ from $\theta$ \cite{cover1999elements}.
Therefore, a larger Fisher information indicates that $\theta$ carries more statistically accessible information about $Z$.

As we show in this Letter, the Fisher information allows quantifying in what order and at what rate the network acquires information about data-generating parameters.
By formulating SGD as a stochastic modified equation (SME), we derive an upper bound on the Fisher information flow from $Z$ to $\theta$.
This bound gives a speed-limit-type constraint on how rapidly information about data parameters can be transferred to the trainable parameters during stochastic learning dynamics.
Specializing to basis-function linear regression, we formulate a mode-resolved analysis, from which characteristic acquisition times and their ordering can be identified.
Moreover, we demonstrate that, under reasonable assumptions, the characteristic acquisition times can be predicted from the speed limit.
Finally, we verify our results in a single-parameter linear model and demonstrate parameter-dependent acquisition time scales for a sinusoidal target learned with Gaussian radial basis functions.

\sectionprl{Stochastic learning dynamics}\label{sec:dynamics}
For concreteness and to fix notation, we consider supervised learning with data generated from a conditional distribution $P(X,Y|Z)$, where $X$ is the input, $Y$ is the output, and $Z$ characterizes the data-generating process.
The training set is $\mathcal{D}=\{(x_i,y_i)\}_{i=1}^N$, with $(x_i,y_i)$ independently sampled from $P(X,Y|Z)$.
The parameters $\theta$ of a model $ f_\theta(x)$ are trained by minimizing a loss function $L_i(\theta)$; for regression, one may take $L_i(\theta) = |y_i - f_\theta(x_i)|^2/2$~\cite{bishop2006pattern}.

In SGD, the parameters are updated at each step $k$ using a mini-batch represented by an index set $\Gamma_k$ of size $m$, whose elements are sampled independently from the set of training indices.
The SGD update rule is given by $\theta_{k+1} = \theta_k - \varepsilon/m \sum_{i \in \Gamma_k} \nabla_\theta L_i(\theta_k)$, where $\varepsilon$ is the learning rate.
We define full-batch drift and the covariance of the mini-batch gradient by $a_Z(\theta) \coloneqq - 1/N \sum_{i=1}^N \nabla_\theta L_i(\theta)$, $D_Z(\theta) \coloneqq \mathrm{Cov}_{\Gamma} \, [- 1/m \sum_{i \in \Gamma} \nabla_\theta L_i(\theta)]$.
For $N\gg m \gg 1$, the mini-batch fluctuation can be approximated as Gaussian, giving the effective dynamics.
\begin{gather} \label{eq:sme}
    \dd \theta_t^{(\chi)} = a_Z(\theta_t^{(\chi)}) \dd t + \sqrt{\chi D_Z(\theta_t^{(\chi)})} \dd W_t,
\end{gather}
where $W_t$ is a Wiener process.
Here, $t=k\varepsilon$ and $\chi \in \{1,\varepsilon\}$ distinguishes two scalings.
The choice $\chi=1$ gives the standard diffusion scaling, whereas $\chi=\varepsilon$ reproduces the one-step noise covariance of mini-batch SGD with learning rate $\varepsilon$ and the latter is called SME or stochastic gradient flow \cite{Li2017-tu,Ali2020-dm}.
The explicit decomposition of SGD into drift and mini-batch noise is given in the End Matter.

We define the \emph{Fisher-information flow} from $Z$ into $\theta$ as
\begin{align} \label{eq:fi-flow}
    \mathcal{J}^{\theta,\chi}_{Z,t}
    \coloneqq
    \begin{cases}
        \displaystyle
        \lim_{\varepsilon \to 0} \frac{\mathcal{F}^{\theta}_{Z,t+\varepsilon} - \mathcal{F}^{\theta}_{Z,t}}{\varepsilon}
        & \chi=1, \vspace{2mm} 
        \\
        \displaystyle
        \mathcal{F}^\theta_{Z,t+\varepsilon} - \mathcal{F}^\theta_{Z,t}
        & \chi=\varepsilon.
    \end{cases}
\end{align}
For a vector-valued $Z$, the following scalar relations may be applied to each component $z$; the corresponding matrix form is obtained by retaining the parameter indices.

\sectionprl{Fisher-information flow speed limit}\label{sec:speed-limit}
The Fisher-information flow obeys the upper bound
\begin{gather} \label{leq:fisl}
    \mathcal{J}_{Z,t}^{\theta,\chi} \leq \mathcal{I}_{Z,t}^{\mathrm{drift},\chi} + \mathcal{I}_{Z,t}^{\mathrm{noise},\chi},
\end{gather}
where we defined
\begin{subequations} \label{eq:total-information-budget}
    \begin{align}
        \mathcal{I}^{\mathrm{drift},\chi}_{Z,t} &\coloneqq \left\langle (\partial_Z a_Z)^\top D_Z^{-1} (\partial_Z a_Z) \right\rangle_t, \\ \mathcal{I}^{\mathrm{noise},\chi}_{Z,t} &\coloneqq \dfrac{\chi}{2\varepsilon} \left\langle \left\|D_Z^{-1/2}(\partial_Z D_Z)D_Z^{-1/2}\right\|_{\mathrm{F}}^2 \right\rangle_t,
    \end{align}
\end{subequations}
which we refer to as the \emph{drift information budget} and \emph{noise information budget}, respectively.
The right-hand side is evaluated at $\theta_t^{(\chi)}$ and averaged over its distribution.

The bound follows directly from the Fisher information of the one-step transition kernel.
Equation~\eqref{eq:sme} gives the Gaussian transition probability $p(\theta_{t+\dd t}|\theta_t,Z) = \mathcal{N}(\theta_{t+\dd t}|\theta_t + a_Z(\theta_t)\dd t, \chi D_Z(\theta_t) \dd t)$.
Its conditional Fisher information is
\begin{equation} \label{eq:cond-fisher-inf}
    \begin{aligned}
        \mathcal{F}^{\theta_{t+\dd t}|\theta_t}_{Z} &\coloneqq \ev{\|\nabla_Z \log p(\theta_{t+\dd t}|\theta_t,Z)\|^2}_t \\
        &= \ev{(\partial_Z a_Z)^\top (\chi D_Z)^{-1}(\partial_Z a_Z)}_t \dd t \\
        &\quad + \dfrac{1}{2} \ev{\left\|D_Z^{-1/2}\bigl(\partial_Z D_Z\bigr)D_Z^{-1/2}\right\|_{\mathrm{F}}^2}_t + o(\varepsilon).
    \end{aligned}
\end{equation}
The chain rule for Fisher information yields $\mathcal{F}^{\theta}_{Z,t+\dd t} - \mathcal{F}^{\theta}_{Z,t} = \mathcal{F}^{\theta_{t+\dd t}|\theta_t}_{Z} - \mathcal{F}^{\theta_t|\theta_{t+\dd t}}_{Z}$~\cite{Zamir1998-wh}.
Because the backward conditional Fisher information $\mathcal{F}^{\theta_t|\theta_{t+\dd t}}_{Z}$ is non-negative, multiplying the chain rule by the normalization appropriate to Eq.~\eqref{eq:fi-flow} and using Eq.~\eqref{eq:cond-fisher-inf} yields Eq.~\eqref{leq:fisl}.

The drift budget is controlled by the sensitivity of the average update direction to $Z$, measured in the inverse-noise metric $D_Z^{-1}$.
Its interpretation is that, while a drift that depends sensitively on the data-generating parameters $Z$ directly transmits this information to the mean value of the network parameters, large fluctuations along a direction suppress the amount of drift information that can be transmitted through that direction.
The noise budget instead quantifies the information about $Z$ carried by the $Z$ dependence of the covariance of the stochastic gradient.
Previous studies have examined how the scale and anisotropic covariance of mini-batch fluctuations affect optimization, minima selection, and generalization \cite{keskar2017iclr-large,le2018iclr-bayesian,pmlr-v97-zhu19e,pmlr-v108-wen20a}.
Here we provide a complementary information-flow perspective: the noise both constrains drift-mediated information transmission and, through its Z-dependent covariance, can itself provide an additional channel of information acquisition.
Importantly, these two quantities form an upper bound rather than an additive decomposition of the actual Fisher-information flow: part of the transition information may be cancelled by the backward conditional term $\mathcal{F}^{\theta_t|\theta_{t+\dd t}}_{Z}$.

\sectionprl{Basis-function linear regression} \label{basis-func-linear-regression}
To obtain more concrete expressions for the Fisher information flow and assess the speed limit, we now specialize the general framework to a model that is linear in its trainable parameters.
We consider $y=g_Z(x)+\eta$, and approximate the target by $f_\theta(x)=\theta^\top \Psi(x)$, $\Psi(x)=(\psi_1(x),\ldots,\psi_d(x))^\top$.
Here, $\eta$ denotes residual noise accounting for statistical errors, and is assumed to be Gaussian with zero mean and variance $\sigma^2$.
Although $f_\theta$ is linear in $\theta$, it can represent a nonlinear function of $x$ through the fixed basis functions $\psi_j$.

Define $H \coloneqq \langle\Psi\Psi^\top\rangle_x$, $r_Z \coloneqq \langle\Psi g_Z\rangle_x$.
Assuming $H$ is nonsingular and using a squared error loss function $L_i = |y_i - f_\theta(x_i)|^2/2$, the population loss $L\coloneqq\sum_{i=1}^NL_i$ can be written as 
\begin{equation}
    \begin{aligned}
        L(\theta) &= \dfrac{1}{2} \ev{(g_Z(x) - \theta^\top \Psi(x))^2}_x + \dfrac{\sigma^2}{2} \\
        &= \dfrac{1}{2} (\theta - \theta_\mathrm{st})^\top H (\theta - \theta_\mathrm{st}) + \mathrm{const}.
    \end{aligned}
\end{equation}
where $\theta_\mathrm{st} = H^{-1} r_Z$ is the optimal parameter.
The drift is therefore $a_Z(\theta) = - H(\theta - \theta_\mathrm{st})$.

Let $\delta_Z(x) \coloneqq g_Z(x) - \theta^\top_\mathrm{st} \Psi(x)$ be the approximation residual.
Near $\theta_\mathrm{st}$, the state dependence of the diffusion matrix is subleading, and its stationary value is $D_\mathrm{st} = \langle(\delta_Z^2 + \sigma^2)\Psi\Psi^\top\rangle_x/m$.
Thus, the near-convergence dynamics reduces to the multivariate Ornstein--Uhlenbeck process
\begin{gather} \label{eq:basis-ou}
    \dd \theta_t^{(\chi)} = - H(\theta_t^{(\chi)} - \theta_\mathrm{st}) \dd t + \sqrt{\chi D_\mathrm{st}} \dd W_t,
\end{gather}
as in related local descriptions of constant-step-size SGD near a quadratic optimum \cite{Mandt2017-fj}.
The detail of derivation, including the terms neglected away from the fixed point, is given in the End Matter.

\sectionprl{Order of information acquisition} \label{sec:order-of-inf-acquisition}
For an initially Gaussian parameter distribution, Eq.~\eqref{eq:basis-ou} preserves Gaussianity.
Its mean and covariance are
\begin{subequations} \label{eq:moment-dynamics}
    \begin{align}
        \mu_t &= (I - \mathrm{e}^{-Ht})\theta_\mathrm{st} + \mathrm{e}^{-Ht} \mu_0, \\
        \Sigma_t^\chi &= \mathrm{e}^{-Ht} \Sigma_0 \mathrm{e}^{-Ht} + \chi \int_0^t \mathrm{e}^{-Hs} D_\mathrm{st} \mathrm{e}^{-Hs} \dd s.
    \end{align}
\end{subequations}
For a scalar component $z$ of $Z$, the Fisher information of a Gaussian distribution is
\begin{equation} \label{eq:gaussian-fisher-inf}
    \begin{aligned}
        \mathcal{F}^{\theta,\chi}_{z,t} &= (\partial_z \mu_t)^\top (\Sigma^\chi_t)^{-1} (\partial_z \mu_t) \\
        &\quad + \dfrac{1}{2} \mathrm{Tr} \qty[(\Sigma^\chi_t)^{-1}(\partial_z\Sigma^\chi_t)(\Sigma^\chi_t)^{-1}(\partial_z\Sigma^\chi_t)].
    \end{aligned}
\end{equation}
In the regimes studied below, the mean contribution of Eq.~\eqref{eq:gaussian-fisher-inf} is dominant.
We therefore neglect the covariance contribution.
Moreover, for $\delta_Z\simeq 0$, the stationary diffusion matrix satisfies $D_\mathrm{st}\simeq\sigma^2H/m$, so that $H$ and $D_\mathrm{st}$ share the same eigenbasis.
Choosing an isotropic initial covariance, $\Sigma_0\propto I$, then makes $\Sigma_0$ diagonal in the same basis.
Thus, $H$, $D_\mathrm{st}$, and $\Sigma_0$ are simultaneously diagonalizable by an orthogonal matrix $U$.
We denote the eigenvalues of $H$ in this basis by $\lambda_i$, the corresponding diagonal elements of $D_\mathrm{st}$ by $\tilde{D}_i$, and those of $\Sigma_0$ by $\Sigma_{0,i}$, such that $U^\top H U = \mathrm{diag} (\lambda_i)$, $U^\top D_\mathrm{st} U = \mathrm{diag} (\tilde{D}_i)$, and $U^\top \Sigma_0 U = \mathrm{diag}(\Sigma_{0,i})$.
The coupling of a data-generating parameter $z$ to these dynamical modes is denoted by $\tilde u_z \coloneqq U^\top \partial_z \theta_\mathrm{st}$, with $\tilde{u}_{z,i}$ its $i$th component.
The Fisher information then decomposes into independent modal contributions,
\begin{gather} \label{eq:modal-fi}
    \mathcal{F}^{\theta,\chi}_{z,t} \simeq \sum_{i=1}^d \dfrac{\tilde{u}_{z,i}^2 (1 - \mathrm{e}^{-\lambda_i t})^2}{\Sigma_{\infty,i}^\chi + (\Sigma_{0,i} - \Sigma_{\infty,i}^\chi \mathrm{e}^{-2\lambda_i t})},
\end{gather}
where $\Sigma_{\infty,i}^\chi=\chi \tilde{D}_i/(2\lambda_i)$.

For $\chi=1$ exactly, and for the SGD scaling $\chi=\varepsilon$ to leading order in $\varepsilon$, the flow associated with mode $i$ is $\mathcal{J}^{\theta,\chi}_{z,i}\simeq\chi\partial_t\mathcal{F}^{\theta,\chi}_{z,i}$.
It reaches its maximum at
\begin{gather} \label{eq:modal-peak-time}
    \tau^\chi_i = \dfrac{1}{\lambda_i} \ln \qty(1 + \sqrt{\dfrac{\Sigma_{0,i}}{\Sigma^\chi_{\infty,i}}}),
\end{gather}
with peak value
\begin{gather} \label{eq:modal-peak-value}
    \max_t \mathcal{J}^{\theta,\chi}_{z,i} = \dfrac{\lambda_i^2 \tilde{u}_{z,i}^2}{\tilde{D}_i} = \mathcal{I}^{\mathrm{drift},\chi}_{z,i}.
\end{gather}
Thus, each data-generating parameter is acquired through the dynamical modes to which $\partial_z \theta_\mathrm{st}$ couples.
When the basis approximation is accurate, $\delta_Z(x)\simeq 0$, one has $\tilde{D}_i=\sigma^2\lambda_i/m$.
Eq.~\eqref{eq:modal-peak-time} then implies that modes with larger $\lambda_i$ are acquired earlier.

Importantly, the peak time of the Fisher-information flow is also reflected in the relaxation of the speed-limit bound.
Near convergence, the transient correction to $\mathcal{I}^{\mathrm{drift},\chi}_{z,i}$ is controlled by the modal mean square deviation $S_{t,i}^\chi\coloneqq \langle [U^\top(\theta_t^{(\chi)} - \theta_\mathrm{st})]_i^2\rangle_t$, which relaxes as 
\begin{gather}
    S_{t,i}^\chi = (S_{0,i} - \Sigma_{\infty,i}) \mathrm{e}^{-2\lambda_i t} + \Sigma_{\infty,i}.
\end{gather}
We define $\tau_{\mathrm{drift},i}^\chi$ as the time at which this decaying transient becomes comparable to the stationary fluctuation scale $\Sigma_{\infty,i}^\chi$.
In the regime $\Sigma_{\infty,i}^\chi \ll \Sigma_{0,i}, S_{0,i}$, both characteristic times reduce to
\begin{gather} \label{eq:time-scales}
    \tau_i^\chi - \tau^\chi_{\mathrm{drift},i} \simeq \dfrac{1}{2\lambda_i} \ln \dfrac{\Sigma_{0,i}}{S_{0,i}}.
\end{gather}
Thus, the speed-limit bound predicts not only the maximal rate of information acquisition but also when this rate reaches its maximum.
This eigenvalue ordering is analogous to spectrum-dependent learning dynamics in linear regression, kernel methods, and linearized neural networks, where modes associated with larger data- or kernel-spectrum eigenvalues are generally learned more rapidly \cite{ADVANI2020428,pmlr-v119-bordelon20a,NEURIPS2018_5a4be1fa}.
Here, however, the ordered quantity is the acquisition of Fisher information about the data-generating parameter $z$, rather than the decay of prediction error.
The full derivation and the conditions underlying Eqs.~\eqref{eq:modal-fi}--\eqref{eq:time-scales} are given in the End Matter.

\sectionprl{Linear target} \label{sec:linear-target}
\begin{figure}[t]
    \centering
    \includegraphics[width=1.0\linewidth]{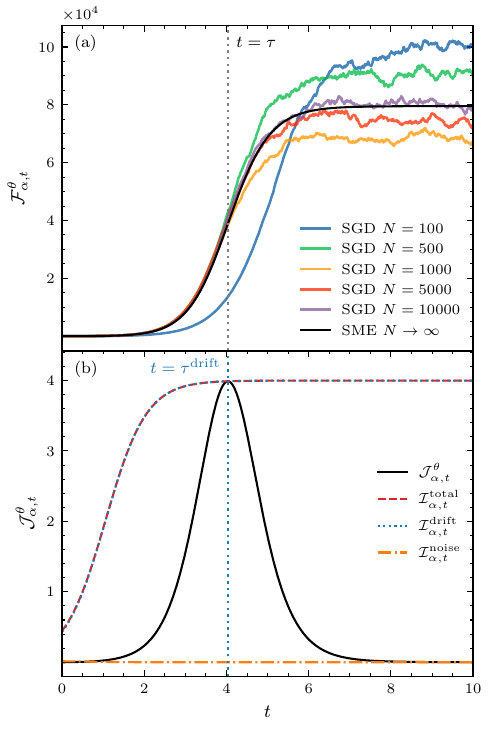}

    \caption{Fisher information dynamics in single-parameter linear regression.
    (a) Time evolution of the Fisher information $\mathcal{F}^\theta_{\alpha,t}$ for SGD with different dataset sizes $N$, together with the SME prediction for $N\to\infty$. The finite-$N$ deviations and fluctuations are reduced for sufficiently large $N$, and the SGD dynamics approaches the SME prediction.
    (b) Fisher-information flow $\mathcal{J}^{\theta}_{\alpha,t}$ (solid line) and its speed-limit bound $\mathcal{I}^\mathrm{total}_{\alpha,t}$, decomposed into drift and noise contributions (dashed lines). The bound is dominated by the drift budget and becomes tight near the peak of $\mathcal{J}^{\theta}_{\alpha,t}$, with nearly identical characteristic times $\tau=4.040$ and $\tau^\mathrm{drift}=4.033$ (dotted lines). 
    Parameters common to both panels are $\alpha=1$, $m=100$, $H=1$, $\sigma=0.5$, and $\varepsilon=0.01$, with the initial parameter distribution $\theta_0\sim\mathcal{N}(0,0.2^2)$.
    }
    \label{fig:linear-fisl}
\end{figure}
We first test the speed limit in a single-parameter model with $y=\alpha x + \eta$, and regression function $f_\theta(x)=\theta x$.
Here, $Z=\alpha$, and $\theta$ acts as a direct estimator of $\alpha$.
From this model, we generate a dataset $\mathcal{D}=\{(x_i,y_i)\}_{i=1}^N$, and assume that $x_i$ follows a normal distribution with zero mean and variance $H$.
For the squared loss, the drift and diffusion coefficients are 
\begin{subequations} \label{eq:linear-sme}
    \begin{align} 
        a_\alpha(\theta) &= - H(\theta - \alpha), \\
        D_\alpha(\theta) &= \dfrac{1}{m} [\sigma^2H + 2H^2(\theta - \alpha)^2].
    \end{align}
\end{subequations}
We set $\chi=\varepsilon$ and evaluate the Fisher information from the mean and variance of an approximately Gaussian parameter distribution.
The resulting drift and noise terms, the numerical differentiation procedure, and the near-convergence analytical solution are given in the End Matter.

Figure \ref{fig:linear-fisl} shows that the Fisher-information flow initially grows, reaches a peak near convergence, and then decreases as the Fisher information approaches its stationary value.
The bound holds throughout the dynamics and becomes tight at the peak.
The analytical Ornstein--Uhlenbeck approximation gives the same peak value as the asymptotic drift budget, $\max_t \mathcal{J}^{\theta}_{\alpha,t}=mH/\sigma^2$, explaining the tightness.
By contrast, the noise budget is appreciable only at an earlier state and does not produce a corresponding feature in the observed flow.
This behavior is consistent with the fact that the loss function in linear regression is convex, so stochastic noise does not play an essential role in efficient exploration of the loss landscape.

\sectionprl{Sinusoidal target} \label{sec:sinusoidal-target}
We next consider $y = A \sin (\omega x + \phi) + \eta, \quad x \sim \mathrm{Unif}[-R,R]$, with data-generating parameters $Z = (A,\omega,\phi)$.
The target is approximated using Gaussian radial basis functions $\psi_j(x) = \exp [- (x-c_j)^2/(2l^2)]$, whose centers $c_j$ are uniformly placed on $[-R,R]$, together with a constant bias basis.
Each basis function is rescaled to $\ev{\psi_j^2}_x=O(1)$.
This construction is linear in $\theta$ while representing a nonlinear target function, and the empirical parameter distributions remain close to Gaussian throughout training.

\begin{figure}[t]
    \centering

    \includegraphics[width=1.0\linewidth]{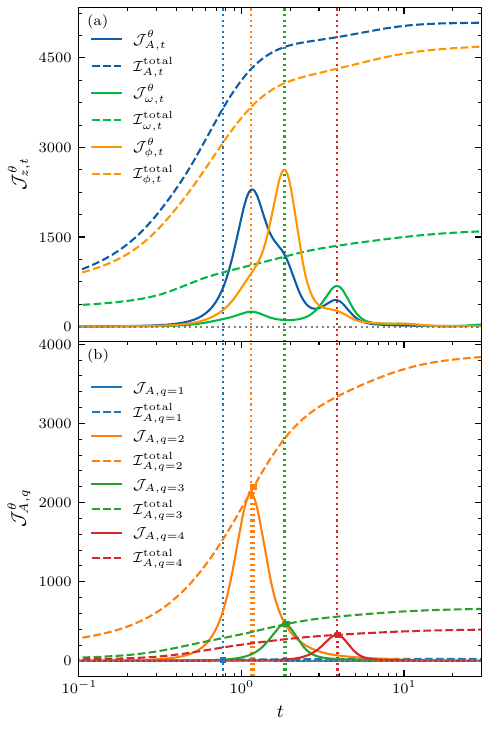}
    
    \caption{Fisher-information flow dynamics in basis-function linear regression for the sinusoidal target $g_Z(x)=A\sin(\omega x + \phi)$.
    (a) Fisher-information flows $\mathcal{J}^{\theta}_{z,t}$ (solid lines) and the corresponding total information budgets $\mathcal{I}^\mathrm{total}_{z,t}$ (dashed lines) for $z=A,\omega,\phi$. The flows exhibit distinct characteristic time scales for different data-generating parameters. Vertical dotted lines indicate the characteristic times $\tau_q$, where $q$ indexes the eigenmodes in descending order of their eivenvalues, $\lambda_1 \geq \lambda_2 \geq \cdots$.
    (b) Mode-resolved Fisher-information flows $\mathcal{J}_{A,q}$ (solid lines) and the corresponding information budgets $\mathcal{I}^\mathrm{total}_{A,q}$ (dashed lines) for $q=1,\ldots,4$. Circular markers indicate the maxima of the modal flows. In contrast to the total flow in (a), each modal flow nearly saturates its corresponding bound at its characteristic time. The target parameters are $A=1$, $\omega=3$, and $\phi=0.4$. The Gaussian radial basis functions are specified by $R=1$, $d=12$, and $l=0.25$.
    }
    \label{fig:basis-function-fisl}
\end{figure}

For each $z \in \{A,\omega,\phi\}$, we compute the Fisher information using Eq.~\eqref{eq:gaussian-fisher-inf} and evaluate the drift and noise budgets.
Figs.~\ref{fig:basis-function-fisl}(a) may appear to show a less tight bound than in the single-parameter case.
This is because the total Fisher-information flow satisfies
\begin{gather} \label{eq:total-modal-bound}
    \max_t \mathcal{J}^{\theta,\chi}_{z,t} \leq \sum_i \mathcal{I}^{\mathrm{drift},\chi}_{z,i},
\end{gather}
since the modal contributions generally reach their maxima at different times.
When resolved into individual eigenmodes, however, each modal flow saturates the corresponding drift information bound at its peak, as shown in Eig.~\ref{fig:basis-function-fisl}(b).
Equality in Eq.~\eqref{eq:total-modal-bound} is attained only when all modal peaks occur simultaneously.
While the characteristic peak time of each mode is set by its eigenvalue $\lambda_i$ under the conditions considered here, the strength with information about a given data-generating parameter $z$ is acquired through that mode is determined by the corresponding coupling $(U^\top\partial_z\theta_\mathrm{st})_i$.

\sectionprl{Discussion}\label{sec:discussion}
In this work, we introduced the Fisher information of the network parameters with respect to data-generating parameters of the training data as a method for analysing information acquisition during the training process.
We found that the corresponding Fisher-information flow can distinguish the rates at which information about different data-generating parameters is encoded in the network and derived a Fisher-information flow speed limit that bounds these rates during stochastic learning dynamics.
For basis-function linear regression, we further identified mode-resolved acquisition times $\tau^\chi_i$, which are determined by the relaxation rates $\lambda_i$ of each mode and its stationary variance $\Sigma_{\infty,i}^\chi$.
Different data-generating parameters couple to these modes with different weights through $\partial_z \theta_\mathrm{st}$, giving rise to distinct information-acquisition dynamics.

Related questions have been studied from the perspective of stochastic thermodynamics, where information processing, prediction, and learning has been connected to entropy production and energetic cost \cite{Still2012-ys,Goldt2017-hg,Goldt2017-lg,wolpert2019stochastic}.
Our result provides a complementary perspective by constraining the rate at which information about parameters of the data-generating process is acquired, rather than the thermodynamic cost of learning.

Fisher information requires identifying the real-valued data-generating parameters underlying the training data and is thus not directly applicable to problems with discrete parameters.
In applications of machine learning to physical problems such as inference of equations of motion \cite{Brunton2016,Bonisegna2018,Champion2019,Greydanus2019,Frishman2020,Seif2021} or optimization of feedback processes \cite{Beeler2021,Kim2021,Whitelam2023,Casert2024,Szamel2026,Whitelam2026}, real-valued data-generating parameters naturally appear in the form of the physical parameters entering the equations of motion.
Since Fisher information is independent of the representation of the network or the definition of the loss function, our approach allows comparing how fast different network architectures and learning strategies acquire information about different physical parameters.

Importantly, the underlying information-theoretic bound is not restricted to SGD, but follows from the Fisher-information chain rule for parameter-dependent Markov transitions.
This suggests potential applications to biological adaptive systems, such as sensory adaptation and synaptic plasticity, where internal states encode information about environmental signals \cite{Hartich2016-zq,Sartori2014-lg,Karbowski2021-ns}.
Applying this framework to such systems could reveal how rapidly information about different environmental properties is acquired and how this acquisition is constrained by the underlying response dynamics and fluctuations.

Although non-Gaussian parameter distributions complicate the analysis for nonlinear neural networks, the inequality itself remains applicable.
For strongly heavy-tailed gradient fluctuations, where L\'{e}vy dynamics may be more appropriate than a diffusion approximation \cite{Simsekli2019-kw}, extending the present framework to jump processes provides a natural direction for future work.
Finally, analysing the Fisher information contained in subsets of trainable parameters may reveal which degrees of freedom encode specific properties of the data-generating process, potentially informing the design of learning architectures.

\section*{Acknowledgments}
S.K. thanks K.Tamano and R.Suzuki for fruitful discussions.
S.K. was supported by JST BOOST, Grant Number JPMJBS2407.
A.D. was supported by JSPS KAKENHI (Grants No. 24H00833 and 25K00926). 
This work was supported by JSPS International Joint Research Program (JRP-LEAD with DFG), grant number 20261606.

\bibliography{refs.bib}

@ARTICLE{Shwartz-Ziv2017-va,
  title         = "Opening the black box of Deep Neural Networks via information",
  author        = "Shwartz-Ziv, Ravid and Tishby, Naftali",
  journal       = "arXiv [cs.LG]",
  month         =  mar,
  year          =  2017,
  archivePrefix = "arXiv",
  primaryClass  = "cs.LG"
}

@INPROCEEDINGS{Li2017-tu,
  title     = "Stochastic modified equations and adaptive stochastic gradient
               algorithms",
  author    = "Li, Qianxiao and Tai, Cheng and E, Weinan",
  editor    = "Precup, Doina and Teh, Yee Whye",
  booktitle = "Proceedings of the 34th International Conference on Machine
               Learning",
  publisher = "PMLR",
  volume    =  70,
  pages     = "2101--2110",
  series    = "Proceedings of Machine Learning Research",
  year      =  2017
}

@INPROCEEDINGS{Ali2020-dm,
  title     = "The Implicit Regularization of Stochastic Gradient Flow for Least Squares",
  author    = "Ali, Alnur and Dobriban, Edgar and Tibshirani, Ryan",
  booktitle = "International Conference on Machine Learning",
  publisher = "PMLR",
  pages     = "233--244",
  month     =  nov,
  year      =  2020
}

@book{cover1999elements,
  title={Elements of information theory},
  author={Cover, Thomas M},
  year={1999},
  publisher={John Wiley \& Sons}
}

@ARTICLE{Weimar2025-rl,
  title     = "Fisher information flow in artificial neural networks",
  author    = "Weimar, Maximilian and Rachbauer, Lukas M and Starshynov, Ilya
               and Faccio, Daniele and Adilova, Linara and Bouchet, Dorian and
               Rotter, Stefan",
  journal   = "Phys. Rev. X.",
  publisher = "American Physical Society (APS)",
  volume    =  15,
  number    =  3,
  month     =  sep,
  year      =  2025
}

@ARTICLE{Horowitz2014-tt,
  title     = "Thermodynamics with Continuous Information Flow",
  author    = "Horowitz, Jordan M and Esposito, Massimiliano",
  journal   = "Phys. Rev. X.",
  publisher = "American Physical Society",
  volume    =  4,
  number    =  3,
  pages     =  031015,
  month     =  jul,
  year      =  2014
}

@ARTICLE{Schreiber2000-nb,
  title     = "Measuring information transfer",
  author    = "Schreiber, T",
  journal   = "Phys. Rev. Lett.",
  publisher = "American Physical Society (APS)",
  volume    =  85,
  number    =  2,
  pages     = "461--464",
  month     =  jul,
  year      =  2000
}

@ARTICLE{Ito2013-ys,
  title     = "Information thermodynamics on causal networks",
  author    = "Ito, Sosuke and Sagawa, Takahiro",
  journal   = "Phys. Rev. Lett.",
  publisher = "American Physical Society (APS)",
  volume    =  111,
  number    =  18,
  pages     =  180603,
  month     =  nov,
  year      =  2013
}

@ARTICLE{Goldt2017-hg,
  title     = "Stochastic thermodynamics of learning",
  author    = "Goldt, Sebastian and Seifert, Udo",
  journal   = "Phys. Rev. Lett.",
  publisher = "American Physical Society (APS)",
  volume    =  118,
  number    =  1,
  pages     =  010601,
  month     =  jan,
  year      =  2017
}

@ARTICLE{Hartich2014-zm,
  title     = "Stochastic thermodynamics of bipartite systems: transfer entropy
               inequalities and a Maxwell’s demon interpretation",
  author    = "Hartich, D and Barato, A C and Seifert, U",
  journal   = "J. Stat. Mech.",
  publisher = "IOP Publishing",
  volume    =  2014,
  number    =  2,
  pages     = "P02016",
  month     =  feb,
  year      =  2014
}

@ARTICLE{Hartich2016-zq,
  title     = "Sensory capacity: An information theoretical measure of the
               performance of a sensor",
  author    = "Hartich, David and Barato, Andre C and Seifert, Udo",
  journal   = "Phys. Rev. E.",
  publisher = "American Physical Society (APS)",
  volume    =  93,
  number    =  2,
  pages     =  022116,
  month     =  feb,
  year      =  2016
}

@book{gardiner2009stochastic,
  title={Stochastic methods},
  author={Gardiner, Crispin},
  volume={4},
  year={2009},
  publisher={Springer Berlin Heidelberg}
}

@book{bishop2006pattern,
  title={Pattern recognition and machine learning},
  author={Bishop, Christopher M and Nasrabadi, Nasser M},
  volume={4},
  number={4},
  year={2006},
  publisher={Springer}
}

@INCOLLECTION{Bottou2010-kh,
  title     = "Large-scale machine learning with stochastic gradient descent",
  author    = "Bottou, Léon",
  booktitle = "Proceedings of COMPSTAT'2010",
  publisher = "Physica-Verlag HD",
  address   = "Heidelberg",
  pages     = "177--186",
  year      =  2010
}

@INPROCEEDINGS{Simsekli2019-kw,
  title     = "A Tail-Index Analysis of Stochastic Gradient Noise in Deep Neural
               Networks",
  author    = "Simsekli, Umut and Sagun, Levent and Gurbuzbalaban, Mert",
  booktitle = "International Conference on Machine Learning",
  publisher = "PMLR",
  pages     = "5827--5837",
  month     =  may,
  year      =  2019,
}

@article{li2019stochastic,
  title={Stochastic modified equations and dynamics of stochastic gradient algorithms i: Mathematical foundations},
  author={Li, Qianxiao and Tai, Cheng and others},
  journal={Journal of Machine Learning Research},
  volume={20},
  number={40},
  pages={1--47},
  year={2019}
}

@inproceedings{keskar2017iclr-large,
  title     = {{On Large-Batch Training for Deep Learning: Generalization Gap and Sharp Minima}},
  author    = {Keskar, Nitish Shirish and Mudigere, Dheevatsa and Nocedal, Jorge and Smelyanskiy, Mikhail and Tang, Ping Tak Peter},
  booktitle = {International Conference on Learning Representations},
  year      = {2017}
}

@ARTICLE{Mandt2017-fj,
  title   = "Stochastic gradient descent as approximate Bayesian inference",
  author  = "Mandt, Stephan and Hoffman, Matthew D and Blei, David M",
  journal = "J. Mach. Learn. Res.",
  volume  =  18,
  number  =  134,
  pages   = "1--35",
  year    =  2017
}

@inproceedings{tishby2015deep,
  title={Deep learning and the information bottleneck principle},
  author={Tishby, Naftali and Zaslavsky, Noga},
  booktitle={2015 ieee information theory workshop (itw)},
  pages={1--5},
  year={2015},
  organization={Ieee}
}

@ARTICLE{Ito2020-es,
  title     = "Stochastic time evolution, information geometry, and the
               Cramér-Rao bound",
  author    = "Ito, Sosuke and Dechant, Andreas",
  journal   = "Phys. Rev. X.",
  publisher = "American Physical Society (APS)",
  volume    =  10,
  number    =  2,
  pages     =  021056,
  month     =  jun,
  year      =  2020
}

@ARTICLE{Shiraishi2018-cz,
  title     = "Speed limit for classical stochastic processes",
  author    = "Shiraishi, Naoto and Funo, Ken and Saito, Keiji",
  journal   = "Phys. Rev. Lett.",
  publisher = "American Physical Society (APS)",
  volume    =  121,
  number    =  7,
  pages     =  070601,
  month     =  aug,
  year      =  2018
}

@ARTICLE{Nishiyama2026-gd,
  title     = "Unified speed limits in classical and quantum dynamics via
               temporal Fisher information",
  author    = "Nishiyama, Tomohiro and Hasegawa, Yoshihiko",
  journal   = "Phys. Rev. E.",
  publisher = "American Physical Society (APS)",
  volume    =  114,
  number    = "1-1",
  pages     =  014120,
  month     =  jul,
  year      =  2026
}

@ARTICLE{Karbowski2024-jb,
  title     = "Bounds on the rates of statistical divergences and mutual
               information via stochastic thermodynamics",
  author    = "Karbowski, Jan",
  journal   = "Phys. Rev. E.",
  publisher = "American Physical Society (APS)",
  volume    =  109,
  number    = "5-1",
  pages     =  054126,
  month     =  may,
  year      =  2024
}

@ARTICLE{Zamir1998-wh,
  title     = "A proof of the Fisher information inequality via a data
               processing argument",
  author    = "Zamir, R",
  journal   = "IEEE Trans. Inf. Theory",
  publisher = "Institute of Electrical and Electronics Engineers (IEEE)",
  volume    =  44,
  number    =  3,
  pages     = "1246--1250",
  month     =  may,
  year      =  1998
}

@ARTICLE{Vo2020-cg,
  title     = "Unified approach to classical speed limit and thermodynamic
               uncertainty relation",
  author    = "Vo, Van Tuan and Van Vu, Tan and Hasegawa, Yoshihiko",
  journal   = "Phys. Rev. E.",
  publisher = "American Physical Society (APS)",
  volume    =  102,
  number    = "6-1",
  pages     =  062132,
  month     =  dec,
  year      =  2020
}

@ARTICLE{Van-Vu2023-ty,
  title     = "Thermodynamic unification of optimal transport: Thermodynamic
               uncertainty relation, minimum dissipation, and thermodynamic
               speed limits",
  author    = "Van Vu, Tan and Saito, Keiji",
  journal   = "Phys. Rev. X.",
  publisher = "American Physical Society (APS)",
  volume    =  13,
  number    =  1,
  pages     =  011013,
  month     =  feb,
  year      =  2023
}

@ARTICLE{Goldt2017-lg,
  title     = "Thermodynamic efficiency of learning a rule in neural networks",
  author    = "Goldt, Sebastian and Seifert, Udo",
  journal   = "New J. Phys.",
  publisher = "IOP Publishing",
  volume    =  19,
  number    =  11,
  pages     =  113001,
  month     =  nov,
  year      =  2017
}

@ARTICLE{Still2012-ys,
  title     = "Thermodynamics of prediction",
  author    = "Still, Susanne and Sivak, David A and Bell, Anthony J and Crooks,
               Gavin E",
  journal   = "Phys. Rev. Lett.",
  publisher = "American Physical Society (APS)",
  volume    =  109,
  number    =  12,
  pages     =  120604,
  month     =  sep,
  year      =  2012
}

@article{wolpert2019stochastic,
  title={The stochastic thermodynamics of computation},
  author={Wolpert, David H},
  journal={Journal of Physics A: Mathematical and Theoretical},
  volume={52},
  number={19},
  pages={193001},
  year={2019},
  publisher={IOP Publishing}
}

@ARTICLE{Tishby2000-nl,
  title         = "The information bottleneck method",
  author        = "Tishby, Naftali and Pereira, Fernando C and Bialek, William",
  journal       = "arXiv [physics.data-an]",
  month         =  apr,
  year          =  2000,
  archivePrefix = "arXiv",
  primaryClass  = "physics.data-an"
}

@ARTICLE{Saxe2019-eb,
  title     = "On the information bottleneck theory of deep learning",
  author    = "Saxe, Andrew M and Bansal, Yamini and Dapello, Joel and Advani,
               Madhu and Kolchinsky, Artemy and Tracey, Brendan D and Cox, David
               D",
  journal   = "J. Stat. Mech.",
  publisher = "IOP Publishing",
  volume    =  2019,
  number    =  12,
  pages     =  124020,
  month     =  dec
}

@article{matsumoto2018role,
  title = {Role of sufficient statistics in stochastic thermodynamics and its implication to sensory adaptation},
  author = {Matsumoto, Takumi and Sagawa, Takahiro},
  journal = {Phys. Rev. E},
  volume = {97},
  issue = {4},
  pages = {042103},
  numpages = {11},
  year = {2018},
  month = {Apr},
  publisher = {American Physical Society}
}

@InProceedings{pmlr-v97-goldfeld19a,
  title = 	 {Estimating Information Flow in Deep Neural Networks},
  author =       {Goldfeld, Ziv and Van Den Berg, Ewout and Greenewald, Kristjan and Melnyk, Igor and Nguyen, Nam and Kingsbury, Brian and Polyanskiy, Yury},
  booktitle = 	 {Proceedings of the 36th International Conference on Machine Learning},
  pages = 	 {2299--2308},
  year = 	 {2019},
  editor = 	 {Chaudhuri, Kamalika and Salakhutdinov, Ruslan},
  volume = 	 {97},
  series = 	 {Proceedings of Machine Learning Research},
  month = 	 {09--15 Jun},
  publisher =    {PMLR}
}

@inproceedings{le2018iclr-bayesian,
  title     = {{A Bayesian Perspective on Generalization and Stochastic Gradient Descent}},
  author    = {Le, Samuel L. Smith and Quoc V.},
  booktitle = {International Conference on Learning Representations},
  year      = {2018}
}

@InProceedings{pmlr-v97-zhu19e,
  title = 	 {The Anisotropic Noise in Stochastic Gradient Descent: Its Behavior of Escaping from Sharp Minima and Regularization Effects},
  author =       {Zhu, Zhanxing and Wu, Jingfeng and Yu, Bing and Wu, Lei and Ma, Jinwen},
  booktitle = 	 {Proceedings of the 36th International Conference on Machine Learning},
  pages = 	 {7654--7663},
  year = 	 {2019},
  editor = 	 {Chaudhuri, Kamalika and Salakhutdinov, Ruslan},
  volume = 	 {97},
  series = 	 {Proceedings of Machine Learning Research},
  month = 	 {09--15 Jun},
  publisher =    {PMLR}
}

@InProceedings{pmlr-v108-wen20a,
  title = 	 {An Empirical Study of Stochastic Gradient Descent with Structured Covariance Noise},
  author =       {Wen, Yeming and Luk, Kevin and Gazeau, Maxime and Zhang, Guodong and Chan, Harris and Ba, Jimmy},
  booktitle = 	 {Proceedings of the Twenty Third International Conference on Artificial Intelligence and Statistics},
  pages = 	 {3621--3631},
  year = 	 {2020},
  editor = 	 {Chiappa, Silvia and Calandra, Roberto},
  volume = 	 {108},
  series = 	 {Proceedings of Machine Learning Research},
  month = 	 {26--28 Aug},
  publisher =    {PMLR}
}

@article{ADVANI2020428,
    title = {High-dimensional dynamics of generalization error in neural networks},
    journal = {Neural Networks},
    volume = {132},
    pages = {428-446},
    year = {2020},
    issn = {0893-6080},
    author = {Madhu S. Advani and Andrew M. Saxe and Haim Sompolinsky}
}

@InProceedings{pmlr-v119-bordelon20a,
  title = 	 {Spectrum Dependent Learning Curves in Kernel Regression and Wide Neural Networks},
  author =       {Bordelon, Blake and Canatar, Abdulkadir and Pehlevan, Cengiz},
  booktitle = 	 {Proceedings of the 37th International Conference on Machine Learning},
  pages = 	 {1024--1034},
  year = 	 {2020},
  editor = 	 {III, Hal Daumé and Singh, Aarti},
  volume = 	 {119},
  series = 	 {Proceedings of Machine Learning Research},
  month = 	 {13--18 Jul},
  publisher =    {PMLR}
}

@inproceedings{NEURIPS2018_5a4be1fa,
 author = {Jacot, Arthur and Gabriel, Franck and Hongler, Clement},
 booktitle = {Advances in Neural Information Processing Systems},
 editor = {S. Bengio and H. Wallach and H. Larochelle and K. Grauman and N. Cesa-Bianchi and R. Garnett},
 pages = {},
 publisher = {Curran Associates, Inc.},
 title = {Neural Tangent Kernel: Convergence and Generalization in Neural Networks},
 volume = {31},
 year = {2018}
}

@INPROCEEDINGS{Yaida2018-qg,
  title     = "Fluctuation-dissipation relations for stochastic gradient descent",
  author    = "Yaida, Sho",
  booktitle = "International Conference on Learning Representations",
  month     =  sep,
  year      =  2018
}

@ARTICLE{Goldt2020-iw,
  title     = "Dynamics of stochastic gradient descent for two-layer neural
               networks in the teacher-student setup",
  author    = "Goldt, Sebastian and Advani, Madhu S and Saxe, Andrew M and
               Krzakala, Florent and Zdeborová, Lenka",
  journal   = "J. Stat. Mech.",
  publisher = "IOP Publishing",
  volume    =  2020,
  number    =  12,
  pages     =  124010,
  month     =  dec,
  year      =  2020,
}

@INPROCEEDINGS{Saad1995-hs,
  title     = "Dynamics of on-line gradient descent learning for multilayer
               neural networks",
  author    = "Saad, D and Solla, S",
  booktitle = "Neural Information Processing Systems",
  pages     = "302--308",
  month     =  nov,
  year      =  1995
}

@INPROCEEDINGS{saxe2014exact,
  title     = "Exact solutions to the nonlinear dynamics of learning in deep linear neural networks",
  author    = "Saxe, Andrew M. and McClelland, James L. and Ganguli, Surya",
  booktitle = "International Conference on Learning Representations",
  year      = 2014
}

@article{JMLR:v22:20-1288,
  author  = {Gerard Ben Arous and Reza Gheissari and Aukosh Jagannath},
  title   = {Online stochastic gradient descent on non-convex losses from high-dimensional inference},
  journal = {Journal of Machine Learning Research},
  year    = {2021},
  volume  = {22},
  number  = {106},
  pages   = {1--51}
}

@INPROCEEDINGS{Abbe2022-ii,
  title     = "The merged-staircase property: a necessary and nearly sufficient
               condition for {SGD} learning of sparse functions on two-layer
               neural networks",
  author    = "Abbe, Emmanuel and Adsera, Enric Boix and Misiakiewicz, Theodor",
  booktitle = "Conference on Learning Theory",
  publisher = "PMLR",
  pages     = "4782--4887",
  month     =  jun,
  year      =  2022
}

@article{JMLR:v25:23-1543,
  author  = {Yatin Dandi and Florent Krzakala and Bruno Loureiro and Luca Pesce and Ludovic Stephan},
  title   = {How Two-Layer Neural Networks Learn, One (Giant) Step at a Time},
  journal = {Journal of Machine Learning Research},
  year    = {2024},
  volume  = {25},
  number  = {349},
  pages   = {1--65}
}

@INPROCEEDINGS{Pensia2018-ah,
  title     = "Generalization error bounds for noisy, iterative algorithms",
  author    = "Pensia, Ankit and Jog, Varun and Loh, Po-Ling",
  booktitle = "2018 IEEE International Symposium on Information Theory (ISIT)",
  publisher = "IEEE",
  pages     = "546--550",
  month     =  jun,
  year      =  2018
}

@INPROCEEDINGS{Futami2023-ok,
  title     = "Time-independent information-theoretic generalization bounds for
               {SGLD}",
  author    = "Futami, Futoshi and Fujisawa, Masahiro",
  editor    = "Oh, A and Naumann, T and Globerson, A and Saenko, K and Hardt, M
               and Levine, S",
  booktitle = "Advances in Neural Information Processing Systems 36",
  publisher = "Neural Information Processing Systems Foundation, Inc. (NeurIPS)",
  address   = "San Diego, California, USA",
  volume    =  36,
  pages     = "8173--8185",
  year      =  2023
}

@ARTICLE{Parsi2024-gy,
  title     = "Stochastic thermodynamics of learning parametric probabilistic
               models",
  author    = "Parsi, Shervin S",
  journal   = "Entropy (Basel)",
  publisher = "MDPI AG",
  volume    =  26,
  number    =  2,
  pages     =  112,
  month     =  jan,
  year      =  2024
}

@ARTICLE{Lang2014-qz,
  title     = "Thermodynamics of statistical inference by cells",
  author    = "Lang, Alex H and Fisher, Charles K and Mora, Thierry and Mehta,
               Pankaj",
  journal   = "Phys. Rev. Lett.",
  publisher = "American Physical Society (APS)",
  volume    =  113,
  number    =  14,
  pages     =  148103,
  month     =  oct,
  year      =  2014
}

@ARTICLE{Sartori2014-lg,
  title     = "Thermodynamic costs of information processing in sensory
               adaptation",
  author    = "Sartori, Pablo and Granger, Léo and Lee, Chiu Fan and Horowitz,
               Jordan M",
  journal   = "PLoS Comput. Biol.",
  publisher = "Public Library of Science (PLoS)",
  volume    =  10,
  number    =  12,
  pages     = "e1003974",
  month     =  dec,
  year      =  2014
}

@ARTICLE{Karbowski2021-ns,
  title     = "Energetics of stochastic {BCM} type synaptic plasticity and
               storing of accurate information",
  author    = "Karbowski, Jan",
  journal   = "J. Comput. Neurosci.",
  publisher = "Springer Science and Business Media LLC",
  volume    =  49,
  number    =  2,
  pages     = "71--106",
  month     =  may,
  year      =  2021
}

@article{Whitelam2023,
  title = {Demon in the Machine: Learning to Extract Work and Absorb Entropy from Fluctuating Nanosystems},
  author = {Whitelam, Stephen},
  journal = {Phys. Rev. X},
  volume = {13},
  issue = {2},
  pages = {021005},
  numpages = {11},
  year = {2023},
  month = {Apr},
  publisher = {American Physical Society}
}

@article{Brunton2016,
  title={Discovering governing equations from data by sparse identification of nonlinear dynamical systems},
  author={Brunton, Steven L and Proctor, Joshua L and Kutz, J Nathan},
  journal={Proceedings of the national academy of sciences},
  volume={113},
  number={15},
  pages={3932--3937},
  year={2016},
  publisher={National Academy of Sciences}
}

@article{Frishman2020,
  title = {Learning Force Fields from Stochastic Trajectories},
  author = {Frishman, Anna and Ronceray, Pierre},
  journal = {Phys. Rev. X},
  volume = {10},
  issue = {2},
  pages = {021009},
  numpages = {32},
  year = {2020},
  month = {Apr},
  publisher = {American Physical Society}
}

@article{Greydanus2019,
  title={Hamiltonian neural networks},
  author={Greydanus, Samuel and Dzamba, Misko and Yosinski, Jason},
  journal={Advances in neural information processing systems},
  volume={32},
  year={2019}
}

@article{Bonisegna2018,
    author = {Boninsegna, Lorenzo and Nüske, Feliks and Clementi, Cecilia},
    title = {Sparse learning of stochastic dynamical equations},
    journal = {The Journal of Chemical Physics},
    volume = {148},
    number = {24},
    pages = {241723},
    year = {2018},
    month = {03},
    issn = {0021-9606}
}

@article{Champion2019,
  title={Data-driven discovery of coordinates and governing equations},
  author={Champion, Kathleen and Lusch, Bethany and Kutz, J Nathan and Brunton, Steven L},
  journal={Proceedings of the National Academy of Sciences},
  volume={116},
  number={45},
  pages={22445--22451},
  year={2019},
  publisher={National Academy of Sciences}
}

@article{Kim2021,
  title={Deep reinforcement learning for feedback control in a collective flashing ratchet},
  author={Kim, Dong-Kyum and Jeong, Hawoong},
  journal={Physical Review Research},
  volume={3},
  number={2},
  pages={L022002},
  year={2021},
  publisher={APS}
}

@article{Beeler2021,
  title = {Optimizing thermodynamic trajectories using evolutionary and gradient-based reinforcement learning},
  author = {Beeler, Chris and Yahorau, Uladzimir and Coles, Rory and Mills, Kyle and Whitelam, Stephen and Tamblyn, Isaac},
  journal = {Phys. Rev. E},
  volume = {104},
  issue = {6},
  pages = {064128},
  numpages = {10},
  year = {2021},
  month = {Dec},
  publisher = {American Physical Society}
}

@article{Casert2024,
  title={Learning protocols for the fast and efficient control of active matter},
  author={Casert, Corneel and Whitelam, Stephen},
  journal={Nature Communications},
  volume={15},
  number={1},
  pages={9128},
  year={2024},
  publisher={Nature Publishing Group UK, London}
}

@article{Szamel2026,
  title = {Machine-learning-derived protocols for information-based work extraction from active particles},
  author = {Szamel, Grzegorz},
  journal = {Phys. Rev. E},
  volume = {113},
  issue = {3},
  pages = {L033402},
  numpages = {5},
  year = {2026},
  month = {Mar},
  publisher = {American Physical Society}
}

@article{Seif2021,
  title={Machine learning the thermodynamic arrow of time},
  author={Seif, Alireza and Hafezi, Mohammad and Jarzynski, Christopher},
  journal={Nature Physics},
  volume={17},
  number={1},
  pages={105--113},
  year={2021},
  publisher={Nature Publishing Group UK, London}
}

@article{Whitelam2026,
  title={A neural-network Maxwell's demon learns cold damping for work extraction},
  author={Whitelam, Stephen and Ciliberto, Sergio and Bellon, Ludovic},
  journal={arXiv preprint arXiv:2607.06822},
  year={2026}
}

\clearpage
\section*{End Matter}
\sectionprl{From mini-batch SGD to the SME}
For comparison, full-batch gradient descent obeys
\begin{gather}
    \theta_{k+1} = \theta_k - \dfrac{\varepsilon}{N} \sum_{i=1}^N \nabla_\theta L_i(\theta_k).
\end{gather}
Adding and subtracting this full-batch gradient in SGD gives
\begin{equation} \label{eq:derivation-sme}
    \begin{aligned}
        \theta_{k+1} - \theta_k &= - \dfrac{\varepsilon}{N} \sum_{i=1}^N \nabla_\theta L_i(\theta_k) \\
        &\quad + \varepsilon \qty[-\dfrac{1}{m} \sum_{i \in \Gamma_k} \nabla_\theta L_i(\theta_k) + \dfrac{1}{N} \sum_{i=1}^N \nabla_\theta L_i(\theta_k)].
    \end{aligned}
\end{equation}
The term in square brackets has zero mean under mini-batch sampling and covariance $D_Z(\theta_k)$.
Approximating it as Gaussian for $N \gg m \gg 1$ and identifying $t=k\varepsilon$ yields Eq.~\eqref{eq:sme}.

For $\chi = \varepsilon$, the covariance accumulated over one interval of duration $\varepsilon$ is $\varepsilon^2 D_Z$, matching the one-step covariance of Eq.~\eqref{eq:derivation-sme} and thus reproducing the natural scaling of SGD.
In contrast, the choice $\chi=1$ corresponds to the conventional diffusion scaling and is often analytically convenient, as it allows standard results for Langevin and Fokker--Planck dynamics to be applied directly~\cite{gardiner2009stochastic}.

\sectionprl{Numerical and Analytical evaluation of Fisher information}
Throughout the numerical analysis, we approximate the parameter distributions as Gaussian and evaluate the Fisher information using Eq.~\eqref{eq:gaussian-fisher-inf}.
The mean and covariance entering this expression are obtained differently for finite-$N$ SGD and for the $N\to\infty$ SME limit.

For finite-$N$ SGD, as shown in Fig.~\ref{fig:linear-fisl}(a), the mean $\mu_t$ and covariance $\Sigma_t$ are estimated from 5000 independent trajectories.
Their derivatives with respect to $\alpha$ are evaluated by centered differences using ensembles generated at $\alpha+\delta\alpha$ and $\alpha-\delta\alpha$, with $\delta\alpha=0.001$.

\begin{figure}[t] 

    \centering
    
    \includegraphics[width=1.0\linewidth]{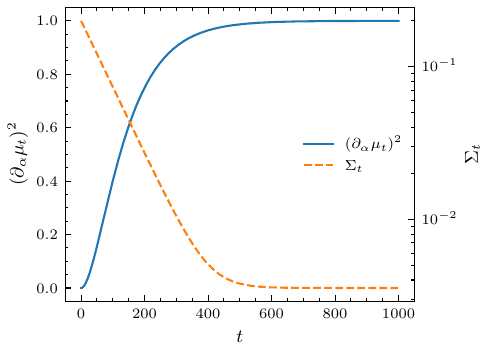}

    \caption{Time evolution of the numerator $(\partial_\alpha\mu_t)^2$ and denominator $\Sigma_t$ of the dominant Fisher information contribution in Eq.~\eqref{eq:end-scalar-fi}. The numerator increases from zero and saturates at unity, whereas the denominator decreases toward the stationary variance set by SGD fluctuations. Their competing time dependences cause the Fisher-information flow to reach a maximum at an intermediate time.}

    \label{fig:analytic-dmu-sigma}

\end{figure}

In the $N\to\infty$ limit, by contrast, the moment dynamics closes under the SME.
The mean $\mu_t$ and covariance $\Sigma_t$ can therefore be obtained analytically from Eq.~\eqref{eq:moment-dynamics}, which apply generally to basis-function linear regression and include the single-parameter linear regression model as a special case.
The Fisher information is then evaluated directly from these analytical moments using Eq.~\eqref{eq:gaussian-fisher-inf}, from which the corresponding Fisher-information flow is obtained.
The results shown in Fig.~\ref{fig:linear-fisl} and Fig.~\ref{fig:basis-function-fisl} are obtained in this manner, without estimating $\mu_t$ or $\Sigma_t$ from stochastic trajectories.
Fig.~\ref{fig:analytic-dmu-sigma} further illustrates the origin of the peak in the Fisher-information flow for the single-parameter linear regression model.

\sectionprl{Analytical form of the speed-limit bound}
The right-hand side of the Fisher-information speed limit can be evaluated directly from the drift and noise budges.
For the single-parameter linear regression model, substituting Eq.~\eqref{eq:linear-sme} into Eq.~\eqref{eq:total-information-budget} gives
\begin{subequations} \label{eq:scalar-total-budgets}
    \begin{align}
        \mathcal{I}^\mathrm{drift}_{\alpha,t} &= \ev{\dfrac{mH}{\sigma^2 + 2H(\theta - \alpha)^2}}_t, \\
        \mathcal{I}^\mathrm{noise}_{\alpha,t} &= \ev{\dfrac{8H^2(\theta-\alpha)^2}{[\sigma^2 + 2H(\theta-\alpha)^2]^2}}_t.
    \end{align}
\end{subequations}
For basis-function linear regression, the corresponding bound can be evaluated analogously by resolving the dynamics into the eigenmodes introduced in the main text.
This mode-resolved representation will be used below in the near-convergence analysis.

\sectionprl{Analytical results near convergence}
\label{end:near-convergence}
We summarize the near-convergence analysis underlying the peak
values and characteristic time scales of the Fisher-information flow.
We first consider the single-parameter linear-regression model.
In the limit $N\to\infty$, the drift and diffusion coefficients are \eqref{eq:linear-sme}
Defining the deviation from the optimum as
$e_t^{(\chi)}\coloneqq\theta_t^{(\chi)}-\alpha$, the condition $(e_t^{(\chi)})^2 \ll \sigma^2/(2H)$ allows us to neglect the state-dependent part of the diffusion coefficient.
The dynamics then reduces to the Ornstein--Uhlenbeck process
\begin{gather}
    \dd e_t^{(\chi)} = - H e_t^{(\chi)}\,\dd t + \sqrt{\chi D_{\mathrm{st}}}\,\dd W_t, \quad D_{\mathrm{st}}=\frac{\sigma^2H}{m}.
    \label{eq:end-scalar-ou}
\end{gather}

For an initially Gaussian distribution independent of $\alpha$, the mean and variance are
\begin{subequations} \label{eq:end-scalar-moments}
    \begin{align}
        \mu_t &= \alpha+(\mu_0-\alpha)\mathrm{e}^{-Ht}, \\
        \Sigma_t^\chi &= \Sigma_\infty^\chi + \bigl(\Sigma_0-\Sigma_\infty^\chi\bigr) \mathrm{e}^{-2Ht},
    \end{align}
\end{subequations}
where $\Sigma_\infty^\chi = \chi \sigma^2/(2m)$.
Since $\partial_\alpha\Sigma_t^\chi=0$, the Gaussian Fisher information reduces to
\begin{gather} \label{eq:end-scalar-fi}
    \mathcal{F}^{\theta,\chi}_{\alpha,t} = \dfrac{ \bigl(1-\mathrm{e}^{-Ht}\bigr)^2}{\Sigma_\infty^\chi + \bigl(\Sigma_0-\Sigma_\infty^\chi\bigr) \mathrm{e}^{-2Ht}}.
\end{gather}
For $\chi=1$ exactly, and for the SGD scaling $\chi=\varepsilon$ to leading order in $\varepsilon$, the corresponding Fisher-information
flow is $\mathcal{J}^{\theta,\chi}_{\alpha,t} \simeq \chi\partial_t\mathcal{F}^{\theta,\chi}_{\alpha,t}$.
Direct differentiation of Eq.~\eqref{eq:end-scalar-fi} gives the peak time
\begin{gather} \label{eq:end-scalar-peak-time}
    \tau^\chi = \dfrac{1}{H} \ln\left(1+\sqrt{\frac{2m\Sigma_0}{\chi\sigma^2}}\right),
\end{gather}
and the peak value
\begin{gather} \label{eq:end-scalar-peak-value}
    \max_t \mathcal{J}^{\theta,\chi}_{\alpha,t} = \frac{\chi H}{2\Sigma_\infty^\chi} = \frac{mH}{\sigma^2}.
\end{gather}
Expanding Eq.~\eqref{eq:scalar-total-budgets} near the optimum yields
\begin{subequations} \label{eq:end-scalar-budget-expansion}
    \begin{align}
        \mathcal{I}^{\mathrm{drift},\chi}_{\alpha,t} &= \dfrac{mH}{\sigma^2} - \dfrac{2mH^2}{\sigma^4} \big\langle(e_t^{(\chi)})^2\big\rangle_t + O\big(\big\langle(e_t^{(\chi)})^4\big\rangle_t\big), \\
        \mathcal{I}^{\mathrm{noise},\chi}_{\alpha,t} &= \dfrac{\chi}{\varepsilon} \dfrac{8H^2}{\sigma^4} \big\langle(e_t^{(\chi)})^2\big\rangle_t + O\big(\big\langle(e_t^{(\chi)})^4\big\rangle_t\big).
    \end{align}
\end{subequations}
Thus, in the small-deviation regime, the right-hand side of the speed limit approaches $mH/\sigma^2$, which coincides with the peak value in
Eq.~\eqref{eq:end-scalar-peak-value}.
This explains why the speed-limit inequality becomes tight near the maximum of the Fisher-information flow.

The same analysis relates the peak time to the relaxation time of the drift budget.
Introducing
\begin{gather}
    S_t^\chi \coloneqq \big\langle(e_t^{(\chi)})^2\big\rangle_t = \Sigma_\infty^\chi + \qty(S_0-\Sigma_\infty^\chi) \mathrm{e}^{-2Ht},
\end{gather}
we define $\tau_{\mathrm{drift}}^\chi$ as the time at which its transient part becomes comparable to the stationary noise level: $(S_0-\Sigma_\infty^\chi) \exp(-2H\tau_{\mathrm{drift}}^\chi) = \Sigma_\infty^\chi$.
This gives
\begin{gather} \label{eq:end-scalar-drift-time}
    \tau_{\mathrm{drift}}^\chi = \dfrac{1}{2H} \ln\qty(\dfrac{S_0-\Sigma_\infty^\chi}{\Sigma_\infty^\chi}).
\end{gather}
When $\Sigma_\infty^\chi\ll\Sigma_0,S_0$,
\begin{gather} \label{eq:end-scalar-time-difference}
    \tau^\chi-\tau_{\mathrm{drift}}^\chi \simeq \frac{1}{2H} \ln\frac{S_0}{\Sigma_0}.
\end{gather}
The two times therefore have the same leading logarithmic dependence on the stationary variance.
At this time, $S_t^\chi=O(\chi\sigma^2/m)$, so the consistency of $(e^{(\chi)})^2\ll\sigma^2/(2H)$ requires $m/\chi\gg H$.

We next consider the basis-function linear regression model.
Defining $e_t^{(\chi)}\coloneqq \theta_t^{(\chi)}-\theta_{\mathrm{st}}$, the drift is $a_Z=-He_t^{(\chi)}$.
The diffusion matrix can be expanded in powers
of $e_t^{(\chi)}$ as
\begin{gather}
    D_Z(\theta_t^{(\chi)}) = D_\mathrm{st} + D_Z^{(1)}(\theta^{(\chi)}_t) + D_Z^{(2)}(\theta_t^{(\chi)}),
\end{gather}
where 
\begin{subequations}
    \begin{align}
        D_\mathrm{st} &\coloneqq \dfrac{1}{m} \langle (\delta_Z^2+\sigma^2) \Psi\Psi^\top \rangle_x, \\
        D_Z^{(1)}(\theta) &\coloneqq -\dfrac{2}{m} \langle\delta_Z(e^\top\Psi)\Psi\Psi^\top\rangle, \\
        D_Z^{(2)}(\theta) &\coloneqq \dfrac{1}{m} \qty[\langle(e^\top\Psi)^2\Psi\Psi^\top\rangle_x - (He)(He)^\top].
    \end{align}
\end{subequations}
Near the optimum, the dynamics is therefore a multivariate Ornstein--Uhlenbeck process given in Eq.~\eqref{eq:basis-ou}.

In this regime, since $\partial_z D_\mathrm{st}=0$, the Fisher information described in Eq.~\eqref{eq:gaussian-fisher-inf} is approximated as
\begin{gather}
    \mathcal{F}^{\theta,\chi}_{z,t} \simeq u_z^\top (I - \mathrm{e}^{-Ht})^\top (\Sigma^\chi_t)^{-1} (I - \mathrm{e}^{-Ht}) u_z,
\end{gather}
where $u_z\coloneqq \partial_z \theta_\mathrm{st}$.
Simultaneously diagonalizing $H$, $D_\mathrm{st}$, and $\Sigma_0$, and using the notation introduced in the main text, this expression reduces to Eq.~\eqref{eq:modal-fi}.
Applying the single-parameter result to each mode yields Eqs.~\eqref{eq:modal-peak-time} and \eqref{eq:modal-peak-value}.
Hence, each modal Fisher-information flow reaches its corresponding
drift information budget at its maximum.

Defining the relaxation time of the drift budget for each mode in the
same manner as in the scalar case gives
\begin{gather}
    \tau_{\mathrm{drift},i}^\chi
    =
    \frac{1}{2\lambda_i}
    \ln\left(
    \frac{
    S_{0,i}-\Sigma_{\infty,i}^\chi
    }{
    \Sigma_{\infty,i}^\chi
    }
    \right).
\end{gather}
For $\Sigma_{\infty,i}^\chi\ll\Sigma_{0,i},S_{0,i}$,
\begin{gather}
    \tau_i^\chi-\tau_{\mathrm{drift},i}^\chi
    \simeq
    \frac{1}{2\lambda_i}
    \ln\frac{\Sigma_{0,i}}{S_{0,i}}.
    \label{eq:end-modal-time-correspondence}
\end{gather}
Thus, the peak of the modal Fisher-information flow and the relaxation of its drift budget are governed by the same characteristic time scale.

Finally, when the basis-function approximation is accurate, $\delta_Z(x)\simeq0$, one has
\begin{gather}
    \widetilde D_i \simeq \frac{\sigma^2\lambda_i}{m}, \qquad \Sigma_{\infty,i}^\chi \simeq \frac{\chi\sigma^2}{2m}.
\end{gather}
If the initial modal variances are comparable, modes with larger $\lambda_i$ therefore peak earlier.
The modal peak times themselves are properties of the learning dynamics, whereas the dependence on the data-generating parameter $z$ enters through the weights $\widetilde u_{z,i}$.
Different data-generating parameters consequently probe different combinations of the modal time scales, giving rise to distinct information-acquisition dynamics.

\clearpage

\setcounter{secnumdepth}{3} 

\setcounter{section}{0}
\setcounter{equation}{0}
\setcounter{footnote}{0}

\renewcommand{\thesection}{S.\Roman{section}}
\renewcommand{\theequation}{S.\arabic{equation}}
\renewcommand{\thefootnote}{*\arabic{footnote}}

\begin{widetext}

\begin{center}
{\large \bf Supplementary Material for \protect \\ ``Speed limits for information acquisition in stochastic learning dynamics''}\\
\vspace*{0.3cm}
Shuta Kobayashi$^{1}$ and Andreas Dechant$^{1}$ \\
\vspace*{0.1cm}
 $^{1}${\small {\it Department of Physics, Kyoto University, Kyoto 606-8502, Japan}}
\end{center}

\section{Analytical calculations near convergence}
\subsection{Single-parameter linear regression for a linear function}
We consider a simple linear regression problem with a linear target function 
\begin{gather}
  y = \alpha x + \eta, \quad x \sim \mathcal{N}(0,H), \quad \eta \sim \mathcal{N}(0,\sigma^2).
\end{gather}
The regression function is given by
\begin{gather}
  f_\theta(x) = \theta x.
\end{gather}
The loss function $L(\theta)$ is then given by
\begin{gather}
  L(\theta) = \dfrac{1}{2N} \sum_{i=1}^N (- (\theta - \alpha) x_i + \eta_i)^2
\end{gather}
and its derivative is
\begin{gather}
  - \nabla_\theta L(\theta) = \dfrac{1}{N} \sum_{i=1}^N (- (\theta - \alpha) x_i + \eta_i) x_i.
\end{gather}

In this case, the drift and diffusion coefficients are respectively given, in the limit $N \to \infty$, by
\begin{subequations}
    \begin{align}
        a_\alpha(\theta) &= \ev{- (\theta - \alpha) x^2 + \eta x}_{x,\eta} = - H (\theta - \alpha), \\
        D_\alpha(\theta) &= \dfrac{1}{m} \mathrm{Var}_{x,\eta} [- (\theta - \alpha) x^2 + \eta x] = \dfrac{\sigma^2 H}{m} + \dfrac{2H^2}{m} (\theta - \alpha)^2.
    \end{align}
\end{subequations}

We define the parameter deviation by
\begin{gather}
  e_t^{(\chi)} \coloneqq \theta_t^{(\chi)} - \alpha.
\end{gather}
Near the convergence point, if
\begin{gather}
    \big(e_t^{(\chi)}\big)^2 \ll \dfrac{\sigma^2}{2H}
\end{gather}
is satisfied, the second-order term can be neglected, and the evolution equation reduces to an Ornstein--Uhlenbeck process:
\begin{equation}
    \begin{aligned}
        \dd \theta_t^{(\chi)} &= - H (\theta_t^{(\chi)} - \alpha) \dd t + \sqrt{\dfrac{\chi \sigma^2H}{m}} \dd W_t \\
        &\equiv a(\theta_t^{(\chi)}) \dd t + \sqrt{\chi D_\mathrm{st}} \dd W_t.
    \end{aligned}
\end{equation}
The evolution equation for the parameter deviation is
\begin{gather}
    \dd e_t^{(\chi)} = -H e_t^{(\chi)} \dd t + \sqrt{\chi D_\mathrm{st}} \dd W_t.
\end{gather}
For $\theta^{\chi}_t$, its mean $\mu_t$ is given by
\begin{gather}
    \big\langle e_t^{(\chi)} \big\rangle = e_0 \mathrm{e}^{-Ht}
\end{gather}
\begin{equation}
    \begin{aligned}
        \therefore \, \mu_t &= (\mu_0 - \alpha) \mathrm{e}^{-Ht} + \alpha \\
        &= (1 - \mathrm{e}^{-Ht}) \alpha + \mu_0 \mathrm{e}^{-Ht}
    \end{aligned}
\end{equation}
The variance $\Sigma_t^{\chi}$ is given by
\begin{equation}
    \begin{aligned}
      \Sigma_t^{\chi} &= \Sigma_0 \mathrm{e}^{-2Ht} + \int_0^t \mathrm{e}^{-2Hs} \chi D_\mathrm{st} \dd s \\
      &= \Sigma_0 \mathrm{e}^{-2Ht} + \dfrac{\chi D_\mathrm{st}}{2H} (1 - \mathrm{e}^{-2Ht}).
    \end{aligned}
\end{equation}
Note that, in the limit $t \to \infty$,
\begin{gather}
  \Sigma_\infty^{\chi} = \dfrac{\chi D_\mathrm{st}}{2H} = \dfrac{\chi \sigma^2}{2m}
\end{gather}
holds.
Under the Gaussian approximation, the Fisher information is given by
\begin{align}
    \mathcal{F}^{\theta,\chi}_{Z,t} = \dfrac{(\partial_\alpha \mu_t)^2}{\Sigma_t^\chi} + \dfrac{1}{2} \dfrac{(\partial_\alpha \Sigma_t^\chi)^2}{(\Sigma_t^\chi)^2}.
\end{align}
Because $\partial_\alpha \Sigma_t^\chi = 0$, the Fisher information reduces to
\begin{gather}
    \mathcal{F}^{\theta,\chi}_{Z,t} = \dfrac{(1 - \mathrm{e}^{-Ht})^2}{\Sigma_\infty^{\chi} + \big(\Sigma_0 - \Sigma_\infty^{\chi}\big)\mathrm{e}^{-2Ht}}
\end{gather}
It therefore follows that the Fisher information ultimately stored in $\theta$ is estimated as
\begin{gather}
    \lim_{t \to \infty} \mathcal{F}^{\theta,\chi}_{Z,t} = \dfrac{1}{\Sigma_\infty^{\chi}} = \dfrac{2m}{\chi \sigma^2}
\end{gather}

We next calculate the Fisher information flow $\mathcal{J}^{\theta,\chi}_{Z,t}$.
Introducing $x\equiv\mathrm{e}^{-Ht},A \equiv \Sigma_0-\Sigma_\infty^{\chi},B \equiv \Sigma_\infty^\chi$, we write
\begin{gather}
    \mathcal{F}(x) = \dfrac{(1-x)^2}{B + Ax^2}
\end{gather}
Its time derivative is
\begin{align}
    \mathcal{J}(x) &= \chi \dv{\mathcal{F}}{x} \dv{x}{t} \\
    &= 2 \chi H \dfrac{x(1-x)(B+Ax)}{(B+Ax^2)^2}.
\end{align}
This derivative corresponds to the Fisher information flow $\mathcal{J}^{\theta,\chi}_{Z,t}$.
The condition $\dd \mathcal{L}/\dd x = 0$ can be written as
\begin{gather}
    \dv{}{x} \qty[\dfrac{x(1-x)(B+Ax)}{(B+Ax^2)^2}] = 0.
\end{gather}
The corresponding point $x_*$ is
\begin{gather}
    x_* = \dfrac{1}{1 + \sqrt{\Sigma_0/\Sigma_\infty^\chi}}
\end{gather}
Therefore, the time $\tau^\chi$ at which $\mathcal{J}^{\theta,\chi}_{Z,t}$ reaches its peak is
\begin{equation}
    \begin{aligned}
        \tau^{\chi} &= \dfrac{1}{H} \log \qty(1 + \sqrt{\dfrac{\Sigma_0}{\Sigma_\infty^{\chi}}}) \\
        &= \dfrac{1}{H} \log \qty(1 + \sqrt{\dfrac{2m\Sigma_0}{\chi \sigma^2}}).
    \end{aligned}
\end{equation}
Under the SGD scaling $\chi=\varepsilon$, the peak shifts to later times as the learning rate $\varepsilon$ decreases.
The peak value is
\begin{gather}
    \max_t \mathcal{J}^{\theta,\chi}_{Z,t} = \left.\mathcal{J}^{\theta,\chi}_{Z,t}\right|_{t=\tau^\chi} = \dfrac{\chi H}{2 \Sigma_\infty^{\chi}} = \dfrac{mH}{\sigma^2}.
\end{gather}

On the other hand, expanding the drift information flow $\mathcal{I}^{\mathrm{drift},\chi}_{Z,t}$ and the noise information flow $\mathcal{I}^{\mathrm{noise},\chi}_{Z,t}$ in powers of the parameter deviation yields
\begin{subequations}
  \begin{align}
    &\begin{aligned}
      \mathcal{I}^\mathrm{drift,\chi}_{\alpha,t} &= \ev{\dfrac{mH}{\sigma^2 + 2H(e_t^{(\chi)})^2}} \\
      &= \dfrac{mH}{\sigma^2} - \dfrac{2mH^2}{\sigma^4} \big\langle (e_t^{(\chi)})^2 \big\rangle + O\big( \big\langle (e_t^{(\chi)})^4 \big\rangle \big),
    \end{aligned} \\
    &\begin{aligned}
      \mathcal{I}^\mathrm{noise, \chi}_{\alpha,t} &= \dfrac{\chi}{\varepsilon} \ev{\dfrac{8H^2(e_t^{(\chi)})^2}{\big(\sigma^2 + 2H (e_t^{(\chi)})^2 \big)}} \\
      &= \dfrac{\chi}{\varepsilon} \dfrac{8H^2}{\sigma^4} \big\langle (e_t^{(\chi)})^2 \big\rangle + O\big(\big\langle (e_t^{(\chi)})^4 \big\rangle \big)
    \end{aligned}
  \end{align}
\end{subequations}
When the parameter deviation is sufficiently small,
\begin{gather}
    \mathcal{I}^{\mathrm{total},\chi}_{Z,t} = \mathcal{I}^\mathrm{drift}_{Z,t} = \dfrac{m H}{\sigma^2}
\end{gather}
which coincides with the peak value of $\mathcal{J}^{\theta,\chi}_{Z,t}$.
Thus, the inequality becomes tight.

We next quantify convergence in order to determine the timescale on which $\mathcal{I}^{\mathrm{drift},\chi}_{Z,t}$ approaches its asymptotic value.
Because the drift information flow was expanded in terms of $S_t^\chi \coloneqq \big\langle (e_t^{(\chi)})^2 \big\rangle$, we regard it as converged once $S_t^\chi$ becomes sufficiently small.
The quantity $S_t^\chi$ is given by
\begin{equation}
  \begin{aligned}
    S_t^\chi &= \Sigma_t - \big\langle e_t^{(\chi)} \big\rangle^2 \\
    &= \qty(S_0 - S_\infty^{\chi}) \mathrm{e}^{-2Ht} + S_\infty^{\chi}.
  \end{aligned}
\end{equation}
We define the convergence time of the drift information flow as the time at which the transient contribution has decayed to the same order as the stationary contribution, namely, when
\begin{gather}
  S^\chi_t - \Sigma^\chi_\infty \simeq \Sigma^\chi_\infty
\end{gather}
is satisfied, and denote it by $\tau_\mathrm{drift}^\chi$.
Substitution gives
\begin{gather}
    (S_0 - \Sigma_\infty^\chi) \mathrm{e}^{-2H \tau^\chi_\mathrm{drift}(c)} = \Sigma_\infty^\chi
\end{gather}
and hence
\begin{align}
    \tau^\chi_\mathrm{drift} &= \dfrac{1}{2H} \ln \qty[\dfrac{S_0 - \Sigma^\chi_\infty}{\Sigma^\chi_\infty}]. \\
\end{align}
When the stationary variance is much smaller than the initial variance, i.e., when $\Sigma^\chi_\infty \ll \Sigma_0$, we obtain
\begin{gather}
    \tau^\chi - \tau^\chi_\mathrm{drift} \simeq \dfrac{1}{2H} \ln \dfrac{S_0}{\Sigma_0}.
\end{gather}
Choosing the initial condition such that $S_0 \simeq \Sigma_0$ gives
\begin{gather}
    \tau^\chi \simeq \tau^\chi_\mathrm{drift}
\end{gather}
and therefore the peak time of the Fisher information flow and the convergence time of the drift information flow approximately coincide, at least to logarithmic order.

This argument additionally requires
\begin{gather}
  S_t^\chi \simeq 2 \Sigma_\infty^\chi = \dfrac{\chi \sigma^2}{m} \ll \dfrac{\sigma^2}{2H}
\end{gather}
Thus, $m/\chi \gg H$ is a necessary condition for the above analytical treatment.

\subsection{Basis-function linear regression for a general target function}
For a more general function $g_Z(x)$, we consider linear regression using basis functions $\psi_j(x)$.
We assume the following data-generating process:
\begin{gather}
    y = g_Z(x) + \eta, \quad \eta \sim \mathcal{N}(0,\sigma^2).
\end{gather}
We define the basis-function vector $\Psi(x)$ as $\Psi = (\psi_1,\psi_2,\ldots,\psi_d)^\top$.
Using a parameter vector $\theta \in \mathbb{R}^d$, the regression function is written as
\begin{gather}
    f_\theta(x) = \theta^\top \Psi(x)
\end{gather}
The loss function is then
\begin{equation}
  \begin{aligned}
      L(\theta) &= \lim_{N \to \infty} \sum_{i=1}^N L_i(\theta) \\
      &= \dfrac{1}{2} \ev{(g_Z(x) - \theta^\top \Psi(x))^2}_x + \dfrac{\sigma^2}{2} \\
      &= \dfrac{1}{2} \qty(\ev{(g_Z(x))^2}_x - 2 \theta^\top \ev{\Psi(x)g_Z(x)}_x + \theta^\top \ev{\Psi\Psi^\top}_x\theta) + \dfrac{\sigma^2}{2} \\
      &= - \dfrac{1}{2} \theta^\top H \theta - \theta^\top r_Z + \dfrac{1}{2} \ev{g_Z^2}_x + \dfrac{\sigma^2}{2},
  \end{aligned}
\end{equation}
where $H \coloneqq \ev{\Psi\Psi^\top}_x,r_Z\coloneqq\ev{\Psi g_Z}_x$.

Let $\theta_\mathrm{st}$ denote the optimal parameter that minimizes the loss function above.
Because we consider a general regression problem, an approximation error that depends on the choice of $\Psi$ may remain even at the optimal parameter.
We define this residual by
\begin{gather}
    \delta_Z(x) = g_Z(x) - \theta_\mathrm{st}^\top \Psi(x).
\end{gather}
The expected squared residual is
\begin{equation}
  \begin{aligned}
      \ev{\delta_Z^2}_x &= \ev{g_Z^2}_x - 2 \theta_\mathrm{st}^\top \ev{\Psi g_Z}_x + \theta_\mathrm{st}^\top \ev{\Psi\Psi^\top}_x \theta_\mathrm{st} \\
      &= \ev{g_Z^2}_x - 2 \theta_\mathrm{st}^\top r_Z + \theta_\mathrm{st}^\top H \theta_\mathrm{st}
  \end{aligned}
\end{equation}
and the loss function can therefore be rewritten as
\begin{align}
    L(\theta) &= \dfrac{1}{2} (\theta - \theta_\mathrm{st})^\top H (\theta - \theta_\mathrm{st}) + \dfrac{1}{2} \ev{\delta_Z^2}_x + \dfrac{\sigma^2}{2}
\end{align}
The gradient is then
\begin{gather}
    \nabla_\theta L(\theta) = H (\theta - \theta_\mathrm{st}).
\end{gather}
Using $e_t^{(\chi)} \coloneqq \theta_t^{(\chi)} - \theta_\mathrm{st}$, the drift term becomes
\begin{gather}
    a_Z(\theta_t^{(\chi)}) = -H e_t^{(\chi)}
\end{gather}

On the other hand, expanding the diffusion coefficient in terms of $e^{(\chi)}_t$ gives
\begin{equation}
  \begin{aligned}
      D_Z(\theta^{(\chi)}_t) &= \dfrac{1}{m} \, \mathrm{Cov}_{x,\eta} \qty[(y(x) - (\theta_t^{(\chi)})^\top \Psi(x))\Psi(x)] \\
      &= \dfrac{1}{m} \, \mathrm{Cov}_{x,\eta} \qty[(g_Z(x) - \theta_\mathrm{st}^\top \Psi(x) - (\theta^{(\chi)}_t - \theta_\mathrm{st})^\top \Psi(x) + \eta) \Psi(x)] \\
      &= \dfrac{1}{m} \, \mathrm{Cov}_{x,\eta} \qty[(\delta_Z(x) - (e^{(\chi)}_t)^\top \Psi(x) + \eta) \Psi(x)] \\
      &= D_\mathrm{st}(\theta^{(\chi)}_t) + D_Z^{(1)}(\theta^{(\chi)}_t) + D_Z^{(2)}(\theta^{(\chi)}_t).
  \end{aligned}
\end{equation}
Here,
\begin{align}
    D_\mathrm{st}(\theta) &= \ev{(\delta_Z^2 + \sigma^2)\Psi\Psi^\top}_x, \\
    D_Z^{(1)}(\theta) &= - \dfrac{2}{m} \ev{\delta_Z(e^\top\Psi)\Psi\Psi^\top}_x, \\
    D_Z^{(2)}(\theta) &= \dfrac{1}{m} \qty[\ev{(e^\top\Psi)^2\Psi\Psi^\top}_x - (He)(He)^\top].
\end{align}
Even when $e_t^{(\chi)}$ is sufficiently small, the diffusion coefficient becomes
\begin{gather}
  D_Z = \dfrac{\sigma^2 H}{m} + \ev{\delta_Z^2(x) \Psi(x)\Psi^\top(x)}_x
\end{gather}

Because this expression is independent of $\theta$, the dynamics near the fixed point again follows an Ornstein--Uhlenbeck process.
The mean $\mu_t$ and covariance $\Sigma_t^\chi$ are then given by
\begin{align}
    \mu_t &= (I - \mathrm{e}^{-Ht})\theta_\mathrm{st} + \mu_0 \mathrm{e}^{-Ht}, \\
    \Sigma_t^\chi &= \mathrm{e}^{-Ht} \Sigma_0 \mathrm{e}^{-Ht} + \int_0^t \mathrm{e^{-Hs}} (\chi D_Z) \mathrm{e}^{-Hs} \dd s
\end{align}
Here, $I$ denotes the $d$-dimensional identity matrix.

Near the convergence point, the Fisher information is approximated as
\begin{equation}
  \begin{aligned}
      \mathcal{F}^{\theta,\chi}_{z,k} &\simeq (\partial_z \mu_t)^\top (\Sigma_t^{\chi})^{-1} (\partial_z \mu_t) \\
      &= u_Z^\top (I - \mathrm{e}^{-Ht})^\top (\Sigma_t^{\chi})^{-1} (I - \mathrm{e}^{-Ht}) u_Z.
  \end{aligned}
\end{equation}
Here, $u_z \coloneqq \partial_z \theta_\mathrm{st}$.
We now assume that $H,D_Z,V_0$ are simultaneously diagonalizable and introduce the eigendecomposition
\begin{align}
    H = U \Lambda U^\top, \quad \Lambda = \mathrm{diag} (\lambda_1,\lambda_2,\ldots,\lambda_d), \\
    \tilde{u} = U^\top u_Z, \quad \widetilde{D} = U^\top D_Z U = \mathrm{diag} (\widetilde{D}_i)
\end{align}
In this basis,
\begin{gather}
    \Sigma_{\infty,i}^{\chi} = \dfrac{\chi \widetilde{D}_i}{2\lambda_i}, \\
    \Sigma_{t,i}^{\chi} = \Sigma_{\infty,i}^{\chi} + (\Sigma_{0,i} - \Sigma_{\infty,i}^{\chi}) \mathrm{e}^{-2\lambda_i t}.
\end{gather}
Therefore,
\begin{gather}
    \mathcal{F}^{\theta,\chi}_{z,t} = \sum_i \dfrac{\tilde{u}_i^2 (1 - \mathrm{e}^{-\lambda_i t})^2}{\Sigma_{\infty,i}^{\chi} + ( \Sigma_{0,i} - \Sigma_{\infty,i}^{\chi}) \mathrm{e}^{-2\lambda_i t}}.
\end{gather}

We define
\begin{gather}
    \mathcal{F}_i(t) = \dfrac{\tilde{u}_i^2 (1 - \mathrm{e}^{-\lambda_i t})^2}{\Sigma_{\infty,i}^{\chi} + (\Sigma_{0,i} - \Sigma_{\infty,i}^{\chi}) \mathrm{e}^{-2\lambda_i t}}
\end{gather}
Defining the Fisher information flow for each mode as $\mathcal{J}_i(t) = \dd_t \mathcal{F}_i(t)$ gives $\mathcal{J}^{\theta,\chi}_{z,t} = \sum_i \mathcal{J}_i(t)$.
As in the scalar case, the maximum value of $\mathcal{J}_i(t)$ and its peak time $\tau_i^\chi$ are obtained as
\begin{gather}
    \max_t \mathcal{J}_i(t) = \mathcal{J}_i(\tau_i^\chi) = \dfrac{\chi \lambda_i \tilde{u}_i^2}{2 \Sigma_{\infty,i}^\chi} = \dfrac{\lambda_i^2 \tilde{u}_i^2}{\widetilde{D}_i} \\
    \tau_i^\chi = \dfrac{1}{\lambda_i} \ln \qty(1 + \sqrt{\dfrac{\Sigma_{0,i}}{\Sigma_{\infty,i}^\chi}})
\end{gather}
The contribution of each mode to the drift information flow is
\begin{gather}
    \mathcal{I}^{\mathrm{drift},\chi}_{z,i} = \dfrac{(\lambda_i \tilde{u}_i)^2}{\widetilde{D}_i}.
\end{gather}
Thus, the peak value coincides with the corresponding modal contribution to the drift information flow.
For the total flow, however,
\begin{gather}
  \mathcal{J}^{\theta,\chi}_{z,t} = \sum_i \mathcal{J}_i(t)
\end{gather}
the peak times generally differ among the modes, and hence
\begin{gather}
  \max_t \mathcal{J}^{\theta,\chi}_{z,t} \leq \sum_i \mathcal{I}^{\mathrm{drift},\chi}_{z,i}
\end{gather}
in general. Equality holds when all modes have the same peak time.

The convergence time of the drift information flow for each mode is
\begin{gather}
    \tau_{\mathrm{drift},i}^\chi = \dfrac{1}{2\lambda_i} \ln \qty(\dfrac{S_{0,i} - \Sigma_{\infty,i}^\chi}{\Sigma_{\infty,i}^\chi})
\end{gather}
and, when $\Sigma_{\infty,i}^\chi \ll S_{0,i}$,
\begin{gather}
    \tau_i^\chi - \tau_{\mathrm{drift},i}^\chi \simeq \dfrac{1}{2\lambda_i} \ln \dfrac{S_{0,i}}{\Sigma_{0,i}}.
\end{gather}
Thus, for each eigenmode, the peak time and the convergence time again have the same characteristic timescale.

When the basis functions are chosen appropriately, $\delta_Z(x) \simeq 0$, and therefore
\begin{gather}
    \widetilde{D} = \dfrac{\sigma^2}{m} U^\top H U = \dfrac{\sigma^2}{m} \Lambda, \\
    \mathrm{i.e.}\,\widetilde{D}_i = \dfrac{\sigma^2 \lambda_i}{m}.
\end{gather}
In this case,
\begin{gather}
  \Sigma_{\infty,i}^{\chi} = \dfrac{\chi}{2\lambda_i} \dfrac{\sigma^2 \lambda_i}{m} = \dfrac{\chi\sigma^2}{2m}
\end{gather}
and the peak time and peak value of each eigenmode are given by
\begin{gather}
    \tau_i^\chi = \dfrac{1}{\lambda_i} \ln \qty(1 + \sqrt{\dfrac{2m\Sigma_0}{\chi \sigma^2}}), \\
    \max_t \mathcal{J}_i(t) = \dfrac{m \lambda_i \tilde{u}_i^2}{\sigma^2}
\end{gather}
Thus, the peaks of the eigenmodes appear in descending order of their eigenvalues.
Conversely, when $\delta_Z(x)$ is large, no such simple relation holds in general.

\end{widetext}

\end{document}